\documentstyle[11pt,paspconf,epsf]{article}

\begin{document}

\title{Observations and Modeling of the Disk-Halo Interaction in our
Galaxy}

\author{Magdalen Normandeau}
\affil{Astronomy Department, 601 Campbell Hall, University of California,
Berkeley, CA, 94720-3411, United States}

\author{Shantanu Basu}
\affil{Canadian Institute for Theoretical Astrophysics, University of
Toronto, 60 St.\ George Street, Toronto, ON, M5S 3H8, Canada}

\begin{abstract}
Galaxies are surrounded by large halos of hot gas which must be
replenished as the gas cools. This led Norman \& Ikeuchi (1989)
to propose the chimney model of the interstellar medium, which
predicts that there should be on the order of a thousand such conduits
connecting the disk and the halo of a galaxy. 

Where then are these structures and other possible disk-halo
connections in our galaxy? What do they look like, how can we detect
them, and what do they tell us about the interstellar medium and about
the Galaxy?

We present a review of the observational evidence for Galactic
disk-halo connections, beginning with large scale searches and then
concentrating on the characteristics of selected candidates. We
summarize how modeling these structures can provide information on
the structure of the interstellar medium in which they evolved,
focusing on the W4 superbubble and the Anchor as illustrations.
\end{abstract}


\def\ga{\mathrel{\hbox{\rlap{\hbox{\lower4pt\hbox{$\sim$}}}\hbox{$>$}}}}

\section{Indications of and predictions for Galactic disk-halo interactions}

\subsection{Why do we need disk-halo interactions?}

In his discussion of disk-halo interactions in external galaxies, Michael 
Dahlem (these proceedings) pointed out that attempting to study
disk-halo interactions in our  
own galaxy leads to the old problem of not being able to see the forest for 
the trees. While it is true that our position within the system prevents us
from gaining a global view, there is much to be learned from close-up 
examinations: the trees have much to tell us about their environment.

Not all galaxies show evidence of disk-halo interaction, in fact not all 
galaxies have a halo as indicated by Dahlem. So what of our
galaxy? Does it have a halo and if so, do the disk and halo interact?
Several observations suggest that they do.

The bulk of the atomic hydrogen (H{\small I}) is near the Galactic plane.
Lockman et al.\ (1986) found two components to its
distribution in the solar neighbourhood: 1)
a Gaussian with $\sigma_z$ = 135 pc (sometimes referred to as the
H{\small I} cloud layer, hereafter the thin H{\small I}
disk), 2) an exponential with a scale height of 500 pc (sometimes
referred to as the H{\small I} intercloud layer, hereafter the
thick H{\small I} disk).
However,
ultraviolet absorption lines of highly ionized species (C{\small IV},
Si{\small IV}, N{\small V}) imply %have evidenced 
the existence 
of a halo of hot gas (greater than at least several $\times 10^4$ K)
with a scale height of $\sim$3 kpc (Savage \& Massa 1987). 
This halo requires 
energy and momentum to maintain it, as well as a source of metals.

Within the halo there are clouds of colder gas, the Intermediate Velocity 
and High Velocity Clouds (IVCs and HVCs; see review by Wakker \& van 
Woerden 1997). These cover more than 10\% of the sky and may contain as much 
as 10\% of the H{\small I} mass of the Galaxy. While some are related to the
Magellanic Stream and others are seen as an Outer Arm extension, a third
group requires some other explanation. In the context of the
chimney model of the Galaxy (Norman \& Ikeuchi 1989), these clouds are 
composed of gas that has cooled after flowing up chimneys; some are still
rising while others are falling back towards the disk. Although metallicity 
determinations remain quite uncertain (by a factor of 3--5) for the HVCs and
IVCs, there have been indications that their heavy element abundances
are comparable to that of ordinary interstellar gas (Danly 1991), 
consistent with the picture of these clouds being cooled processed gas from
the plane.

Closer to home, the so-called Reynolds layer (see the contribution by 
Reynolds in these proceedings) of warm, ionized
gas has a scale height of $\sim$1 kpc but its source of ionization is 
unclear. A significant fraction ($\sim$15\%, MacLow in these proceedings)
of the ionizing photons from the Galactic O stars would suffice but these
stars are almost exclusively confined to the disk and it is therefore difficult
for their ionizing radiation to escape to higher latitudes. Chimneys could
be the solution to this quandary by providing conduits through which
the photons 
could travel unimpeded away from the disk. In addition, Norman (1991)
suggested that the
walls contribute to the ionization of 
upper Galactic layers through diffuse, re-emitted radiation, and that
this would affect a much wider angular range than does the 
radiation escaping directly up the conduit.

A final element which was called upon by Norman \& Ikeuchi in support of
their model was the filling factor of the Hot Ionized Medium (HIM) in the disk,
$\leq$20\% at the solar circle (Ferri\`ere, these proceedings)
which is substantially lower than predicted in the McKee \& Ostriker (1977)
model of the ISM. By confining the hot gas to chimney conduits and evacuating
it to higher latitudes, a lower disk HIM filling factor is obtained.

\subsection{How many superbubbles and chimneys should there be?}

Estimates of the number of chimneys in our galaxy vary considerably. 
%MN: Again, Galaxy capitalized above
Norman \& Ikeuchi (1989) estimated
that, for a steady state, there should be $1\,000$ such conduits. However,
this is an estimate of the total number of superbubbles, based on 
the rate of type II supernovae, the fraction of early type stars belonging
to OB associations, and the expected number of supernovae in a single 
OB association. The assumption is then implicitly made that all superbubbles
are chimneys. This is however unrealistic. Not all superbubbles will blow
out of the disk and into the halo. In fact, the
magnetic field of the Galaxy may prevent most superbubbles from blowing out
(Tomisaka 1990, 1998).

A more reasonable estimate is the empirical one by Heiles et al.\ (1996).
Based on an extrapolation of the worms observed in one quadrant, they
predicted the existence of at least 50 and probably no more than 100
worms in our 
galaxy. However, worms are not necessarily chimneys, as explained below.

\section{Where are the Galactic chimneys?}

\subsection{Wide field searches: Worms}

While cataloguing H{\small I} shells and supershells, Heiles (1984) pointed out
the presence of ``wiggly gas filaments crawling away from the Galactic
plane''. These he dubbed worms and postulated that they may be
remnants of supershells that have opened at the top. 

Following this introductory work, 
Koo et al.\ (1992) set about drawing up an inventory of worm
candidates. To produce
their list of 118 structures,
they made use of two existing low resolution H{\small I} surveys
(Weaver \& Williams 1973; Kerr et al.\ 1986), some supplementary 
H{\small I} observations at the Hat Creek Radio Observatory, and the IRAS
60 $\mu$m and 100 $\mu$m images. They integrated the H{\small I} over the
entire velocity range (--200 to +200 km/s) then applied a median
filter to this image as well as to the infrared ones. An ``object''
present in all three was considered a worm candidate. While this
method provides an objective sample of possible worms, the integration
over all velocities will mask some true worms and possibly create
false ones, and clearly many of the candidates are not proper worms because
they are not perpendicular to the plane. 

A second, less objective yet possibly less misleading catalogue was
produced by Heiles et al.\ (1996) and contains twenty-seven worms.
Their selection was based on morphology at 2695 MHz (data from Reich et al.\
1990) and Radio Recombination Lines (RRLs). 
All of their worms are within the first
quadrant of the Galaxy ($\ell_{\rm max}$ = 61.5\deg) and twelve of the fifteen
for which distances can be evaluated are within the Galactocentric azimuth 
range $90\deg < \theta < 180\deg$; this is the origin of the estimate
mentioned in the previous section.

\subsection{A few examples}

\label{sec:obs_examples}

With these catalogues of potential chimneys and superbubbles available, one
would think that many of these structures would have been studied in detail.
One would be mistaken. Only a very few structures possibly denoting a disk-halo
interaction have been closely examined. 

\subsubsection{The Stockert Thermal Spur}

The Stockert Thermal Spur, or Stockert Chimney, was the first chimney
candidate to receive attention (M\"{u}ller et al.\ 1987). The structure 
originally studied was a spur extending from 2\deg\ to 8\deg\ in latitude,
above the S54 H{\small II} region. Its spectral index indicated thermal
emission, consistent with the chimney picture in which the walls are
photoionized. At a kinematic distance of 2.9 kpc, this structure is 300
pc high, putting its tip above the thin H{\small I} disk. 

While the position and velocity of S54 suggest a relationship between it
and the spur, the stars in the H{\small II} region cannot have caused outflow
up to 300 pc (the outflow velocity would far too great, $\sim$4\,500 km/s) 
nor can they account for 
the ionization. 

This structure is associated with a worm in the Heiles et al.\ (1996) catalogue
(GW18.5+2.8). These authors point out that if this is the worm in its entirety 
then it is curious in that it is not limb brightened. Furthermore they detected
RRL emission from $\ell \approx 15\deg$ to 21\deg and suggested that
the thermal 
spur is in fact only one wall of a chimney or superbubble and that it is 
powered by the M16 cluster which is centred at the base of the region of RRL
emission.

\subsubsection{The W4 superbubble/chimney}

\begin{figure}[t]
\vspace{7cm}
\includegraphics{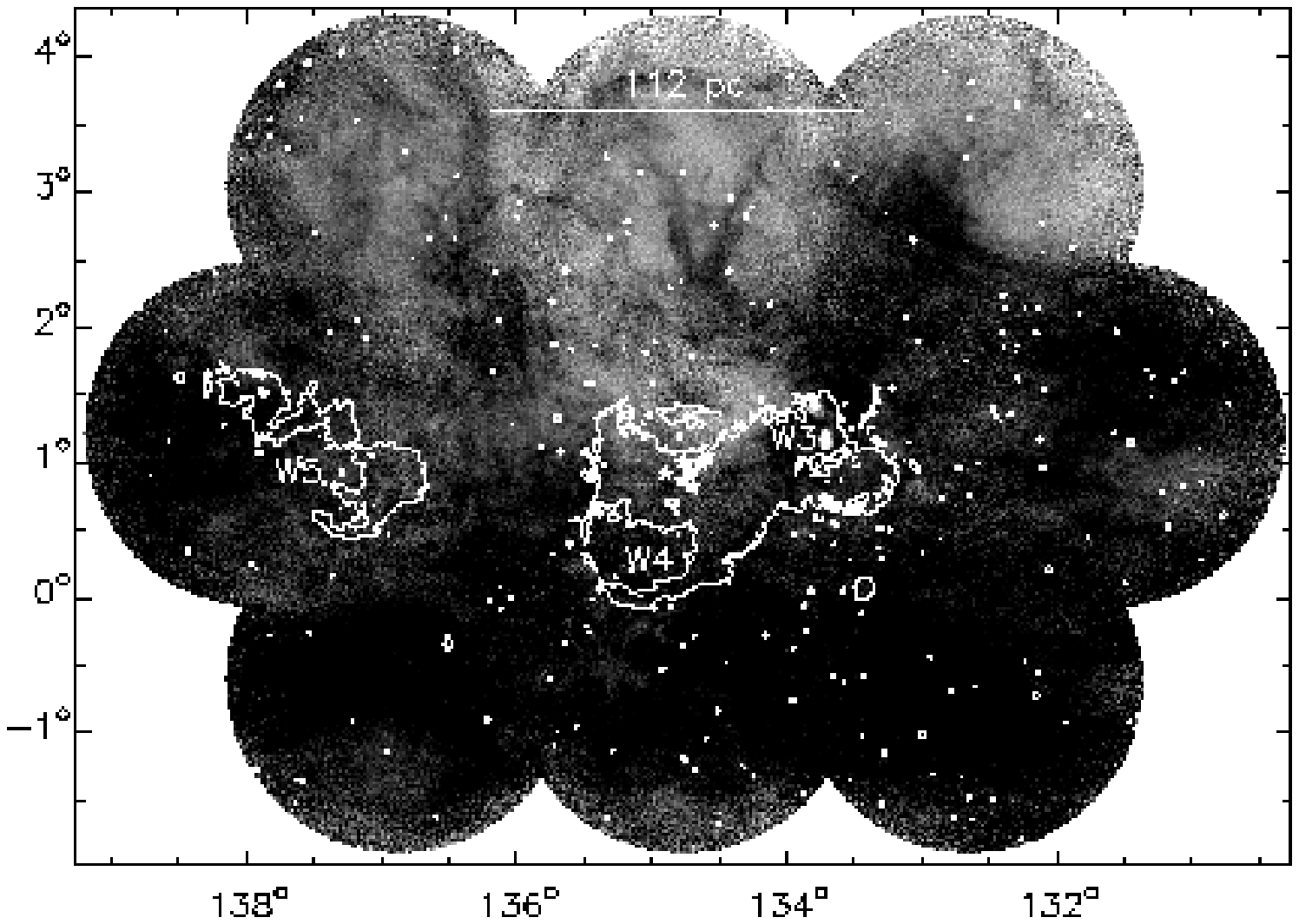}
\caption{The W4 superbubble/chimney}{The greyscales show the H{\small I}
emission (white$\leq$0 K, black$\geq$80 K) at --43.4 km/s and the contours
outline the radio 
continuum emission at 1420 MHz (5 K). The W3, W4, and W5 H{\small II}
regions are labeled and the star symbols indicate the positions of the
O-stars in the OCl 352 cluster.
The scale given at the top of the cone assumes a distance of 2.35 kpc
(Massey et al.\ 1995). These data are from the CGPS pilot project
(Normandeau et al.\ 1997).}
\label{fig:w4_pilot}
\end{figure}

While the composite structure encompassing the Stockert Thermal Spur
is suggestive 
of a superbubble which has burst, there is no indication of outflow towards
higher Galactic latitude. The Canadian Galactic Plane Survey (CGPS) pilot 
project revealed a conical void in the H{\small I}
distribution above the W4 H{\small II} region (Normandeau et al.\ 1996),
within which there are features suggestive of such an outflow 
(Figure~\ref{fig:w4_pilot}).
The V-shaped streamers seen within the chimney at --45.0 km/s are the
culmination 
of a development from a compact cloud at the latitude of the base of the
V at $v_{\rm LSR} = -33.5$ km/s which gradually extends as velocity becomes
more negative.

The walls have an inner lining which is visible in the infrared as well as
in the radio continuum, with a spectral index indicating thermal emission.
There is an H{\small I} arc at $b \sim 3.8\deg$ but it does not completely
close off the shell and it was originally postulated that this is a
remnant of the supershell that had burst to create the chimney. Recent
observations have shown that the eastern wall, as seen in H{\small I}, 
extends to a latitude slightly greater than 6\deg\ (Normandeau 1998). 

The energy source powering this structure seems clear: at the base of the
cone lies the OCl 352 cluster which contains nine O-stars, one of which is 
O4 and two are O5. The wind luminosity of these stars can certainly account for
the evacuation of the cavity seen in the pilot data and their age is in 
agreement with the time required for the cavity to expand to its
current size (see discussion in \S\ref{sec:mod_W4}), as well
as allowing enough 
time for the streamers to have stretched to their present length. 
%However it remains doubtful that the stellar winds are blowing all the 
%way to the halo (see \S\ref{sec:mod_W4}).

H$\alpha$ observations by Dennison et al.\ (1997) suggest the presence of
a cap at $b \sim 7\deg$, corresponding to a height of approximately 200 pc
above the star cluster. While their detection is marginal and there is no
evidence for closure of the supershell in the extended H{\small I}
observations, this is not an impossible or even an improbable situation as
shall be explained below when modeling is discussed. It suggests
that while the W4 superbubble has certainly broken through the thin
disk of atomic hydrogen, it has not broken out of the thick disk to connect
to the halo. However, it has reached into the Reynolds layer and can
contribute to its ionization.

\subsubsection{The Anchor}

\begin{figure}[t]
\vspace{10cm}
\includegraphics{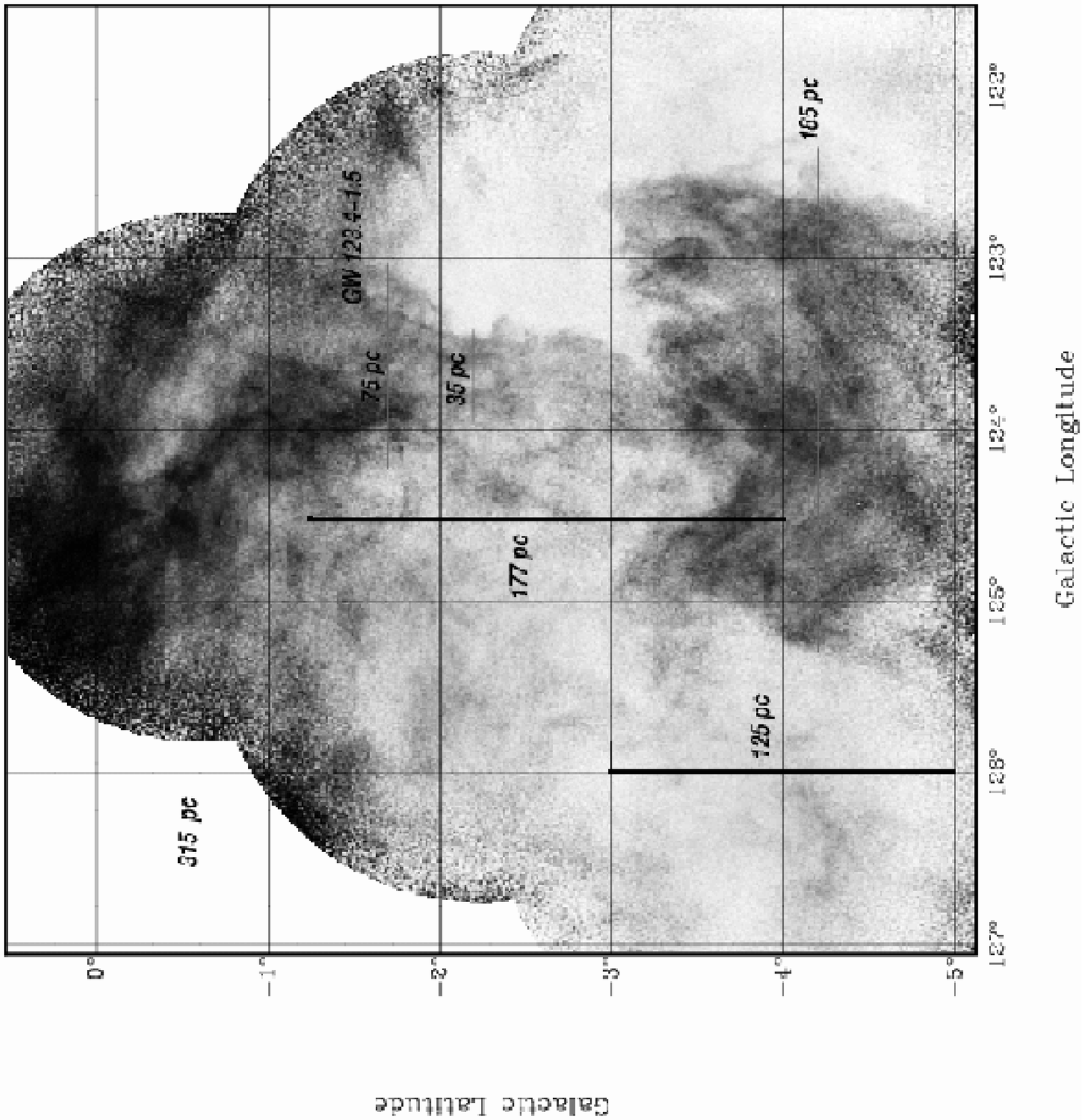}
\caption{The Anchor}{Data from DRAO observations in the H{\small I}
spectral line. Darker shadings correspond to brighter H{\small I}
emission. The scales indicated are for a distance of 3.6 kpc. Figure
courtesy Jayanne English.} 
\label{fig:anchor}
\end{figure}

More recently, the Galactic worm GW123.4--1.5 has been receiving attention
(English et al.\ 1999). This anchor-shaped H{\small I} feature appears to
be dangling from the Galactic plane at a longitude of 124\deg\  
(Figure~\ref{fig:anchor}). 

The stem extends 2.8\deg\ perpendicular to the plane of the Galaxy, and is
topped by a 2.9\deg\ wide cap. The structure covers a total velocity width 
of $\sim$27 km/s, with the cap being redshifted from the stem by $\sim$5 km/s.
The stem appears to be hollow and expanding radially.
With a central velocity of --41 km/s the anchor is at a kinematic distance 
of 3.6 kpc. This would imply a stem length of 177 pc and the top of the cap 
would be $> 300$ pc from the plane. 
The scales indicate that once again
we are looking at a structure connecting the thin disk and thick disk, rather
than the disk and the halo.

The energy source of the Anchor remains a mystery. While the latitude
of the H{\small II} region S185 places it at the base of the stem, its radial 
velocity is approximately --3 km/s, inconsistent with the velocities
of the Anchor.  

%The scales indicate that once again
%we are looking at a structure connecting the thin disk and thick disk, rather
%than the disk and the halo. Nonetheless modeling of this structure is
%instructive and shall be discussed below.

\subsection{Where do we go from here?}

Although disk-halo interactions in the form of chimneys would elegantly explain 
several observed Galactic phenomena, we have yet to see a clear-cut example 
of such a conduit, though we have begun examining structures connecting the
thin disk and the layers directly above it.

The search for chimneys continues. Several avenues are possible. 
Catalogues of worms using various data sets should and are being compiled
(e.g.,\ from the Leiden-Dwingeloo H{\small I} survey, Burton \&
Hartmann 1994). Follow-up observations
of worms and worm candidates should be carried out at higher resolution in the
radio continuum, recombination lines, and H{\small I} spectral line. 
In addition, since
the interior of a chimney is occupied by hot gas which
is being evacuated to the halo, good candidates should be examined for
X-ray emission and absorption lines of highly-ionized species.

\section{Models of disk-halo interactions}
\label{sec:models}

There are several models in the literature which can be applied to a 
study of observed disk-halo interaction candidates. Some of these
models are reviewed below since their general features can
be used to interpret observations and gain insight into the 
disk-halo candidates and their environment.

The earliest blowout model is by Kompaneets (1960),
who found an analytic solution for the shape of a blast wave propagating
into an exponentially stratified atmosphere. This model simply calculates
the shape of a strong shock propagating into a pressure-free environment,
without taking into account the inertia of the swept up mass.
Although initially motivated
to study interactions with the Earth's atmosphere, it can and has
been applied in an astrophysical context as well. 
More sophisticated models of bubble expansion in an astrophysical context
include the thin-shell approximation (MacLow \& McCray 1988),
which determines the expansion speed of the bubble through 
numerical integration of the momentum equation
for various segments of the thin shell of swept-up gas.
This approach accounts for the inertia of the swept-up shell,
and external pressure and gravity can also be included.
Finally, full numerical integration of the hydrodynamic equations
(Tomisaka \& Ikeuchi 1986; MacLow et al.\ 1989; Tenorio-Tagle et al. 1990) 
and magnetohydrodynamic equations (Tomisaka 1990, 1998) yield
the most complete solutions to date. 

Though there are some differences between the various models (see Komljenovic,
Basu \& Johnstone, these proceedings), all the models 
reveal the following general scenario.
The bubble maintains a near-spherical expansion while its radius is
less than or equal to the atmospheric scale height $H$, and if it
expands beyond this height, it begins a rapid acceleration in the
(vertical) direction of stratification, while continuing a decelerating
expansion in the lateral direction. By late times, and at the time of
blowout, the ratio of radius at source height to atmospheric scale height
$R(z=0)/H \approx 2$. The Kompaneets model also predicts that the ratio
of maximum radius to scale height $R_{\rm max}/H \approx 3$ at late times.

Although the models with finite external pressure do show that blowout
does not always occur (see MacLow \& McCray 1988 for a blowout condition),
the models alone cannot tell us whether or not blowout should be common
in our galaxy. This is because the blowout condition depends on the
scale height of interstellar gas (MacLow \& McCray 1988) and the 
scale height of the interstellar magnetic field (Tomisaka 1998). A large
scale height component in either quantity can effectively confine most
superbubbles, but since these parameters (especially the magnetic field) 
are not well constrained observationally at high latitudes, 
it is uncertain whether blowout is common. 

Further insight into the blowout process is best obtained by comparing
models with the observed structures, as these may give us insight
into their ambient environment and thereby whether blowout may 
be commonplace.

\section{Modeling the observed structures}

\label{sec:mod_examples}

The W4 superbubble and the Anchor are the two structures connecting the 
thin disk of H{\small I} to the layers above for which the most
detailed information is now available. With these two objects, we
can take the first steps away from the generic models described in 
\S~3, toward models evolving in more specific environments.
Such modeling can yield information about the interstellar medium
and also highlight constraints on the formation of disk-halo interactions.

\subsection{W4 superbubble}

\label{sec:mod_W4}

The conical shape of the H{\small I} cavity (Figure~\ref{fig:w4_pilot}) provides
an ideal application  
for models of superbubble expansion in a stratified (but smoothly varying)
atmosphere. The observed shape of the cavity bears many similarities to
those predicted by the various models discussed in \S~3.

As mentioned earlier, the H{\small I} maps suggest that the cavity 
is open on top, 
i.e., it is a chimney. However, the H$\alpha$ map of Dennison et al. (1997),
which extends to $b \sim 8\deg$, reveals a shell of H$\alpha$ emission 
(presumably the swept up shell surrounding the cavity that is illuminated 
by the O stars) that reaches a maximum diameter at $b \sim 4\deg$ and 
becomes narrower above, apparently closing at $b \sim 7\deg$.

Recent modeling of the W4 superbubble by Basu, Johnstone \& Martin (1999,
hereafter BJM) reveals that an open cavity in H{\small I} and a closed shell
in H$\alpha$ are mutually consistent, as discussed below. 
BJM fit the shape of the cavity and H$\alpha$ shell
with an analytic Kompaneets profile, yielding straightforward estimates
of the atmospheric structure and bubble age. The dependence of the solution
on various parameters is most transparent when using this model.
Figure 3 shows the H$\alpha$ map of Dennison et al. (1997) overlaid with
the best fit Kompaneets profile. The unmistakable narrowing of the H$\alpha$ 
shell diameter above $b \sim 4\deg$ means that blowout has not yet occurred,
according to any of the theoretical models described in \S~3.

\begin{figure}
\vspace{8cm}
\includegraphics{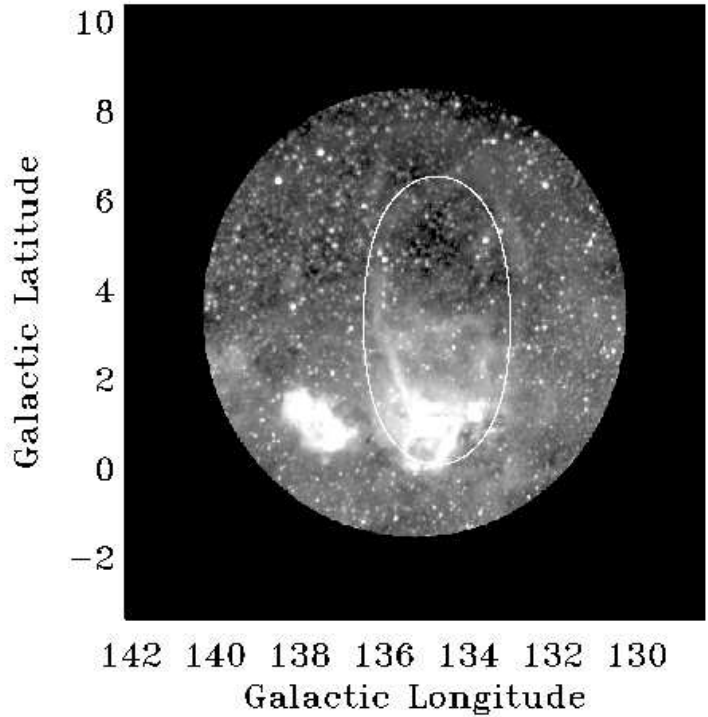}
\caption{The H$\alpha$ map of Dennison et al. (1997) overlaid with the
best fit Kompaneets profile (from BJM).}
\end{figure}

The W4 superbubble is highly elongated, implying that it 
has already expanded through significant vertical stratification.
The radius of the bubble near the cluster must be approximately two
scale heights, as discussed earlier.
By matching the model to the observation, BJM demonstrated 
the unavoidable consequence that $H \approx 25$ pc near W4. This is based
on the distance estimate $d=2.35$ kpc to the OCl 352 cluster (Massey
et al. 1995), which is similar to various previous estimates of $d \approx
2$ kpc to the star cluster and H{\small II} regions.

How are we to interpret such a low value of $H$, in comparison to 
estimates $H \ga 100$ pc for the mean scale height of the thin H{\small I} 
disk in our galaxy? The answer probably lies in the fact that W3/W4/W5 is 
one of the major star-forming complexes in the outer Galaxy, where significant
vertical compression of the interstellar gas must have taken place.
The superbubble has sampled the distribution of molecular and cold H{\small I}
gas near the cluster. On the other hand, we also note that the H$\alpha$
shell extends to some 240 pc above the cluster, and it is remarkable that
the shell maintains its oval shape over such a distance. Various models
of bubble expansion (e.g., MacLow \& McCray 1988) have shown that a 
bubble changes shape dramatically when it travels from a relatively low
scale height atmosphere (e.g., the thin H{\small I} disk) to one with 
greater scale height (e.g., the thick H{\small I} disk); that is, the 
bubble radius expands to become comparable to the local scale height. This
has not happened to the W4 superbubble up to a height $z \approx 240$ pc.
In fact, Komljenovic, Basu \& Johnstone (these proceedings) argue
that the bubble is so highly collimated that even a single atmosphere
hydrodynamic model cannot adequately fit its shape, although the 
simpler Kompaneets model can. They argue that a significant vertical 
component of the magnetic field can be collimating the upper portion 
of the bubble, and that this may also explain the apparent lack of a 
Rayleigh-Taylor instability in the upper H$\alpha$ shell, which is presumably
accelerating. 
However, this does not significantly change the scale height estimate
given above. 

\begin{figure}
\vspace{7cm}
\includegraphics{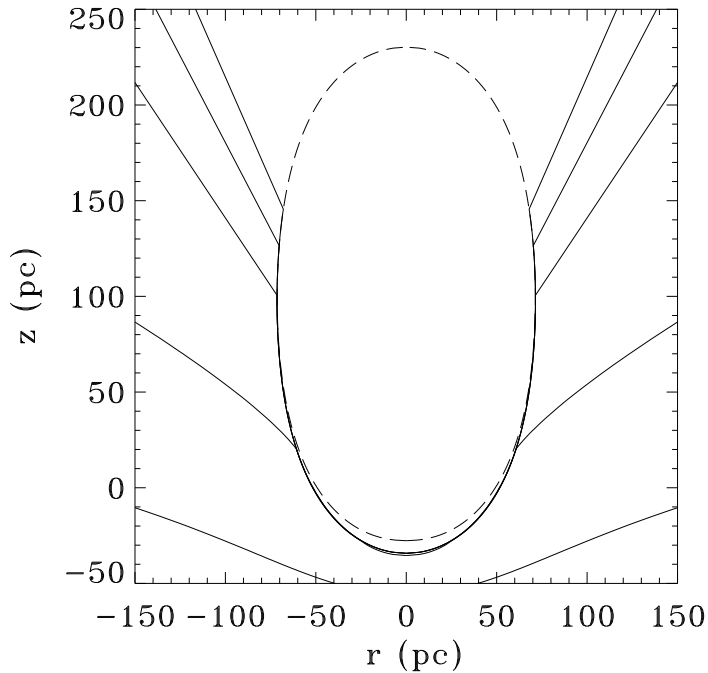}
\caption{Ionization front (solid line) around the best-fit
Kompaneets model for W4 (inner wall of shell in dashed line). The bubble
is embedded in an exponential atmosphere
$n(z) = n_0 \exp(-z/H)$. Parameters used are $\Phi_* = 2.3 \times 10^{50}$
s$^{-1}$ and $H=25$ pc, while $n_0 =$ (bottom to top) 
1, 5, 10, 15, 20 cm$^{-3}$ (from BJM).} 
\end{figure}

In addition to the powerful stellar winds which drive the W4 superbubble,
the nine O stars in the OCl 352 cluster also produce an extremely strong
ultraviolet radiation field, with a flux of Lyman continuum photons
$\Phi_* \simeq 2.3 \times 10^{50}$ s$^{-1}$. This flux encounters a 
highly stratified atmosphere, so the resulting ionization front does
not have the simple shape of a Str\"{o}mgren sphere. BJM have modeled the shape
of the ionization front resulting from the interaction of this ionizing 
flux with an exponentially stratified medium 
which has an embedded cavity and
swept-up shell of matter as predicted by the best fit Kompaneets model. 
Figure 4 shows the location of the ionization front for various choices
of the mean density $n_0$ near the cluster. In all cases, the ionization
front opens up in a cone-like manner at some height, meaning that 
ionizing photons can escape to the Galactic halo within the cone. 
This is due to the low column of matter in the upper portion of the 
shell, since the diverging streamlines of an expanding superbubble 
continually push matter to the sides and cannot transport much matter
to large heights (see MacLow, these proceedings). 
The breakout of ionizing photons near the top explains why one
will not observe neutral H{\small I} above the bubble, giving it 
the appearance of a chimney. This despite the fact that an
upper shell exists and can be 
distinguished from its surroundings in H$\alpha$ emission.
The curves in Figure 4 place the constraint that $n_0 \geq 5$ cm$^{-3}$ 
in the W4 region, since the ionization front is observed to be bounded
at the latitude of the cluster.
Furthermore, the observed drop-off in H{\small I} emission
at $z \approx 100$ pc is best fit by the $n_0 = 10$ cm$^{-3}$ curve,
so BJM adopt this as the most likely value.
These values are in good agreement with the observational
estimate $n_0 \sim 5$ cm$^{-3}$ at the latitude of the cluster 
(Normandeau et al. 1996).
Incidentally, in this model about 15\% of the Lyman continuum photons
from the cluster escape through the top of the shell and can ionize
layers above the thin disk. Fortuitously, this is the same as the
estimated 15\% of Galactic O-star 
ionizing photons that is believed necessary to account for the ionization
of the Reynolds layer (see contributions by MacLow and Reynolds).

It is interesting to note that although BJM's estimates for $n_0$ and
$H$ in the W4 region are considerably higher and lower, respectively, 
than the mean ISM values, the column density $n_0H$ is only slightly higher
than the corresponding mean ISM value.
By obtaining estimates for $n_0$ and $H$, and using an observational estimate
for the wind luminosity $L_0$, BJM also found the
age of the superbubble: $t \approx 2.5$ Myr. This is in agreement with 
various age estimates for the cluster, and supports the idea that the 
superbubble
is blown by stellar winds in a cluster which is too young to have experienced
any supernovae.

\subsection{The Anchor}

Although initially classified as a ``worm'', the Anchor is clearly
not just a superbubble wall, but an object in its own right. A preliminary
estimate of its kinetic energy, based on its velocity width and estimated mass,
is $\sim 2 \times 10^{50}$ ergs (English et al. 1999). Since
these kinds of energies are most readily supplied by supernovae or 
stellar winds, it is natural to wonder whether the Anchor is another
superbubble. 

The unusual mushroom shape of the object defies a simple explanation,
unlike the W4 superbubble which has the conical shape
expected from most models.
In particular, the extreme contrast between a narrow stem and wide cap
is difficult to explain in the context of superbubble models.
However, a stem plus cap morphology can be produced in some circumstances
(see the models of Tenorio-Tagle et al.\ 1990). In particular, it
requires a sharp break from a stratified atmosphere to an effectively
constant density atmosphere. The stem is then the cylindrical cavity
created by an effective ``blowout'' from the stratified atmosphere, and
the cap is the result of quasi-spherical expansion when the hot gas reaches
the uniform density ``halo''. This kind of model was originally used to
represent a true disk-halo interaction, but if the Anchor is a superbubble,
then the interaction is occurring only a few hundred pc away from the Galactic
plane. Perhaps it can represent the interaction between the thin disk and
thick disk components of H{\small I}. However, the superbubble interpretation
leads to the following conclusions: the radius of the stem must be 
approximately two local scale heights, and the bottom of the cap must 
correspond to the height at which the sharp break in the atmosphere occurs.
The distance to the Anchor, as estimated from kinematics is $d = 3.6 \pm
1$ kpc (English et al. 1999), so using even the widest point of the 
stem yields a scale height
$H \approx 20 \pm 6$ pc and ``halo'' (or thick disk) height $z \approx
170 \pm 47$ pc. 

The above numbers show that the superbubble hypothesis requires
a rather unusual atmospheric structure near the Galactic plane.
Yet another
concern with the superbubble hypothesis is the following: the Anchor is
characterized by an apparent {\it excess} of H{\small I} emission relative 
to the background,
rather than the cavity in H{\small I} (as with W4) that is expected to
be occupied 
by hot $\sim 10^6$ K gas. The final word on this issue will await further
analysis of observations, as there is some indication that the stem {\it is}
hollow, and that the cap is redshifted out of the velocity interval
of the Perseus arm. The latter could give the Anchor the appearance of 
being in a 
relatively empty region, when in fact the ambient gas would be displaced into
different velocity channels (English et al. 1999).
 
The Anchor bears a striking resemblance to a thermal plume, e.g.,\ the shape of
a rising fireball after a nuclear explosion. Processes such as rising plumes
or jets may need to be considered, although their origin in the
ISM with energy $\ga 10^{50}$ ergs remains a mystery. An advantage
of such processes is that they can more readily transport matter vertically,
as the Anchor appears to be doing. In contrast, superbubbles are very 
inefficient at transporting matter upwards, as discussed earlier.

\section{Conclusion}

Observations point to the existence of a hot halo around our galaxy, yet its
origin and the means by which it is maintained have not been conclusively
determined. Disk-halo interactions such as chimneys seem likely and could 
account for the presence of hot coronal gas and HVCs within the halo, as 
well as allow the ionization of the Reynolds layer above the thin
H{\small I} disk. A
chimney model of the ISM would also account for the low filling factor of the
HIM in the Galactic disk.

Observations to date have not shown structures clearly connecting the disk
and the halo. The observed worms as well as the Stockert Thermal Spur, 
W4 superbubble, and the Anchor 
are all confined to within a few hundred parsecs from the Galactic plane,
hence they probably connect the thin and thick components of
Galactic H{\small I}, but there is no evidence that they extend all the way 
to the halo.
However, the information that these structures yield about the ISM at low
latitudes can have implications for the disk-halo relationship.
If the cold gas in star-forming regions is as strongly stratified near the 
disk as implied by the W4 superbubble (and also the Anchor if it is indeed
a superbubble), the hot gas in the bubbles will be efficiently channeled
upwards, yielding a relatively low filling factor for the HIM
in the disk (recall that the radius of the hot gas bubble in the disk does
not exceed two local scale heights). 
The elongated cavities which break out of the thin H{\small I} disk
will also allow a significant fraction
of ionizing photons to escape upwards and contribute to the Reynolds layer.

\acknowledgments

We gratefully acknowledge Jayanne English who provided information
about, and pretty pictures of, the Anchor in advance of publication.

\end{document}